\documentclass{emulateapj}
\usepackage{colordvi}
\usepackage{epsfig}
\usepackage{graphicx}
\usepackage{amssymb}
\begin{document}

\title{CANDELS: The progenitors of compact quiescent galaxies at
  $z$$\sim$2}

\author{Guillermo Barro\altaffilmark{1},
S.~ M.~ Faber\altaffilmark{1},
Pablo G.~P\'{e}rez-Gonz\'{a}lez\altaffilmark{2,3},
David C. Koo\altaffilmark{1}, 
Christina C. Williams\altaffilmark{4}, 
Dale D.~Kocevski\altaffilmark{1}, 
Jonathan R.~Trump\altaffilmark{1},
Mark Mozena\altaffilmark{1}, 
Elizabeth McGrath\altaffilmark{1}, 
Arjen van der Wel\altaffilmark{5},
Stijn Wuyts\altaffilmark{6},
Eric F.~ Bell\altaffilmark{7}, 
Darren J.~ Croton\altaffilmark{8}, 
Avishai Dekel\altaffilmark{9},
M.~L.~N.~ Ashby\altaffilmark{10},
Henry C. Ferguson\altaffilmark{11}, 
Adriano Fontana\altaffilmark{12},
Mauro Giavalisco\altaffilmark{4}, 
Norman A. Grogin\altaffilmark{11}, 
Yicheng Guo\altaffilmark{4}, 
Nimish P. Hathi\altaffilmark{13}, 
Philip F. Hopkins\altaffilmark{14},
Kuang-Han Huang\altaffilmark{11}, 
Anton M. Koekemoer\altaffilmark{11}, 
Jeyhan S. Kartaltepe\altaffilmark{15}, 
Kyoung-Soo Lee\altaffilmark{16},
Jeffrey A. Newman\altaffilmark{17}, 
Lauren A. Porter\altaffilmark{1},
Joel R. Primack\altaffilmark{1},
Russell E.~ Ryan\altaffilmark{11}, 
David Rosario\altaffilmark{6},
Rachel S. Somerville\altaffilmark{18}}

\altaffiltext{1}{University of California, Santa Cruz}
\altaffiltext{2}{Universidad Complutense de Madrid}
\altaffiltext{3}{Steward Observatory, University of Arizona}
\altaffiltext{4}{University of Massachusetts}
\altaffiltext{5}{Max-Planck-Institut f\"{u}r Astronomie}
\altaffiltext{6}{Max-Planck-Institut f\"{u}r extraterrestrische Physik}
\altaffiltext{7}{Department of Astronomy, University of Michigan}
\altaffiltext{8}{Swinburne University of Technology}
\altaffiltext{9}{The Hebrew University}
\altaffiltext{10}{Harvard-Smithsonian Center for Astrophysics}
\altaffiltext{11}{Space Telescope Science Institute}
\altaffiltext{12}{INAF Osservatorio Astronomico di Roma}
\altaffiltext{13}{Observatories of the Carnegie Institution of Washington}
\altaffiltext{14}{University of California Berkeley}
\altaffiltext{15}{National Optical Astronomy Observatory}
\altaffiltext{16}{Purdue University}
\altaffiltext{17}{University of Pittsburgh}
\altaffiltext{18}{Rutgers University}
\slugcomment{Submitted to the Astrophysical Journal Letters} 
\begin{abstract}  

We combine high-resolution {\it HST}/WFC3 images with multi-wavelength
photometry to track the evolution of structure and activity of massive
($M_{\star}>10^{10}M_{\odot}$) galaxies at redshifts $z=1.4-3$ in two
fields of the Cosmic Assembly Near-infrared Deep Extragalactic Legacy
Survey (CANDELS). We detect compact, star-forming galaxies (cSFGs)
whose number densities, masses, sizes, and star formation rates
qualify them as likely progenitors of compact, quiescent, massive
galaxies (cQGs) at $z=1.5-3$. At $z\gtrsim2$, most cSFGs have specific
star-formation rates (sSFR$\sim10^{-9}$yr$^{-1}$) half that of
typical, massive SFGs at the same epoch, and host X-ray luminous AGNs
30 times ($\sim$30\%) more frequently. These properties suggest that
cSFGs are formed by gas-rich processes (mergers or disk-instabilities)
that induce a compact starburst and feed an AGN, which, in turn,
quench the star formation on dynamical timescales (few
10$^{8}$yr). The cSFGs are continuously being formed at $z=2-3$ and
fade to cQGs down to $z\sim1.5$. After this epoch, cSFGs are rare,
thereby truncating the formation of new cQGs. Meanwhile, down to
$z=1$, existing cQGs continue to enlarge to match local QGs in size,
while less-gas-rich mergers and other secular mechanisms shepherd
(larger) SFGs as later arrivals to the red sequence. In summary, we
propose two evolutionary tracks of QG formation: an early
($z\gtrsim2$), fast-formation path of rapidly-quenched cSFGs fading
into cQGs that later enlarge within the quiescent phase, and a slow,
late-arrival ($z\lesssim2$) path in which larger SFGs form extended
QGs without passing through a compact state.

\end{abstract}
\keywords{galaxies: starburst --- galaxies: photometry --- galaxies:
  high-redshift}

\section{Introduction}\label{intro}

Nearby galaxies come in two flavors \citep{kauffman03}: red quiescent
galaxies (QGs) with old stellar populations, and blue young
star-forming galaxies (SFGs).  This color bimodality seems to be
already in place at $z\sim2-3$ (\citealt{2010ApJ...709..644I};
\citealt{brammer11}), presenting also strong correlations with mass,
size and morphology: SFGs are typically larger than QGs of the same
mass (\citealt{williams10}; \citealt{wuyts11b}) and disk-like, whereas
QGs are typically spheroids characterized by concentrated light
profiles \citep{bell11}.  Since SFGs are the progenitors of QGs, their
very-different, mass-size relations restrict viable formation
mechanisms.

\begin{figure*}
\centering \includegraphics[width=8.7cm,angle=0.]{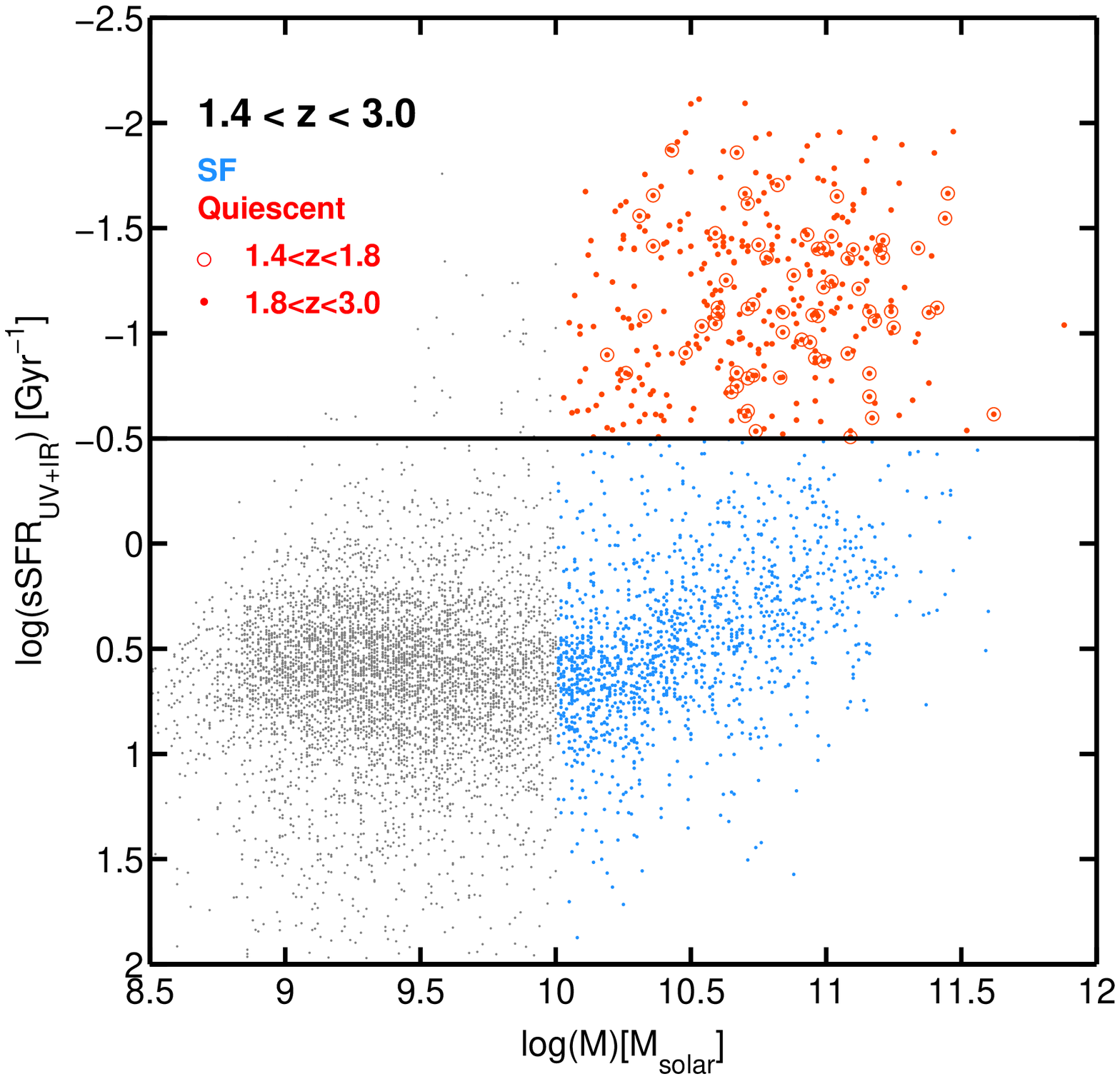}
\includegraphics[width=8.4cm,angle=0.]{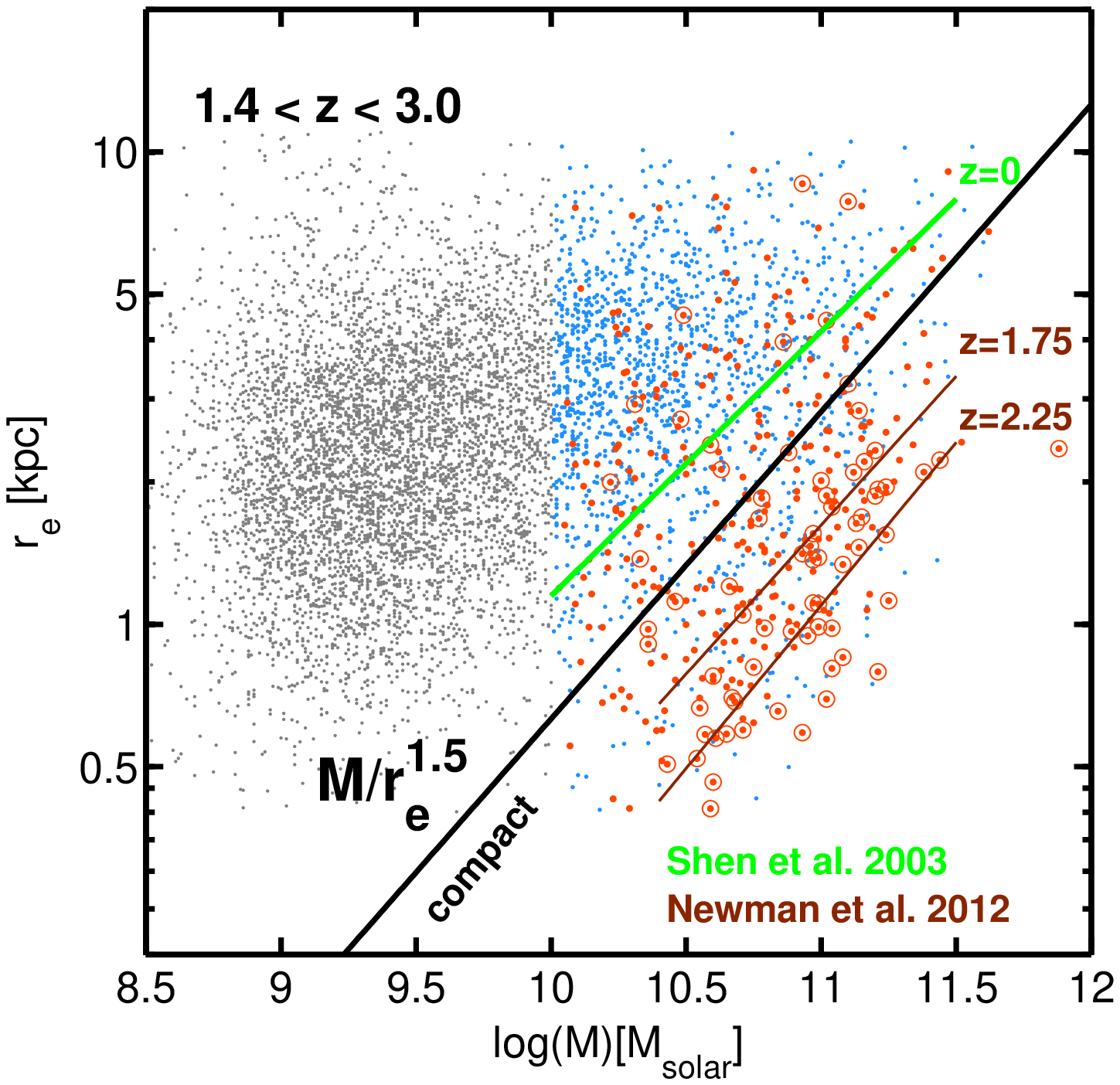}
\caption{\label{ssfr_size} {\it Left panel:} Specific SFR as a
  function of the stellar mass for galaxies at $1.4<z<3.0$. The solid
  black line defines our threshold, $\log(\mathrm{sSFR})=-0.5$, to
  select QGs (red in both panels) and SFGs (blue) above
  $M_{\star}>10^{10}M_{\odot}$.  Grey dots show galaxies with stellar
  masses below the mass selection limit. {\it Right panel:} Stellar
  mass-size relation at $1.4<z<3.0$. The solid black line defines our
  selection criterion for compact galaxies,
  $M/r$$_{e}^{1.5}$$\equiv$$\Sigma_{1.5}$$=10.3M_{\odot}$kpc$^{-1.5}$. The
  green line shows the local mass-size relation for elliptical
  galaxies \citep{shen03}.  The thin brown lines are the mass-size
  relations for QGs found at $z=1.75$ and $z=2.25$ by
  \citet{newman12}.}
\end{figure*}

A major surprise has been the discovery of smaller sizes for massive
QGs at higher redshifts -- these compact QGs (cQGs), also colloquially
known as ``red nuggets", are $\sim5$ times smaller than local,
equal-mass analogs (\citealt{2007MNRAS.382..109T};
\citealt{cassata11}; \citealt{szo11}). In contrast, most of the
massive SFGs at these redshifts are still relatively large disks
(\citealt{kriek09}).  We adopt the view that galaxy mass growth is
accompanied by size growth, as suggested by the mass-size relation.
In this case, to form compact QGs from SFGs, three changes are
required: a significant shrinkage in radius, an increase in mass
concentration, and a rapid truncation of the star formation.

Proposed mechanisms to create compact spheroids from star-forming
progenitors generally involve violent, dynamical processes
(\citealt{naab07}), such as gas-rich mergers \citep{hopkins06} or
dynamical instabilities fed by cold streams \citep{dekel09}. Recent
hydrodynamical simulations of mergers have reproduced some of the
observed properties of cQGs \citep{wuyts10}, if high amounts of cold
gas, as observed by \citet{tacconi10}, are adopted.

If cQGs are so formed, we expect to see a co-existing population of
compact SFGs and recently-quenched galaxies at $z\gtrsim2$. Recent
works demonstrate the existence of such populations
(\citealt{cava03}; \citealt{wuyts11b}; \citealt{whitaker12}), but a direct evolutionary
link has not yet been clearly established.
 
This letter shows a quantitative connection between cSFGs and QGs at
high-z. We combine the deepest photometric data from the optical to
the far IR from the Great Observatories Origins Deep Survey (GOODS;
\citealt{goods}), the UKIDSS Ultra Deep Survey (UDS), the Cosmic
Assembly Near-infrared Deep Extragalactic Legacy Survey (CANDELS;
\citealt{candelsgro}; \citealt{candelskoe}),
FIDEL\footnote{\anchor{http://irsa.ipac.caltech.edu/data/SPITZER/FIDEL/}{http://irsa.ipac.caltech.edu/data/SPITZER/FIDEL/}},
and
SpUDS\footnote{\anchor{http://irsa.ipac.caltech.edu/data/SPITZER/SpUDS/}{http://irsa.ipac.caltech.edu/data/SPITZER/SpUDS/}}
to estimate stellar masses, SFRs, and sizes for massive, high-z
galaxies. By analyzing the global evolution in the space defined by
these parameters, we suggest two paths (fast and slow) for QG
formation from $z\sim3$ to $z\sim1$.

We adopt a flat cosmology with $\Omega_{M}$=0.3,
$\Omega_{\Lambda}$=0.7 and H$_{0}=70$~km~s$^{-1}$~Mpc$^{-1}$.
 
\section{Data Description and sample selection}

This letter is based on a sample of massive galaxies built from the
{\it HST}/WFC3 F160W ($H$-band) selected catalog
($H_{5\sigma}(\mathrm{AB})=27$~mag) for the GOODS-S and UDS fields of
CANDELS.  Consistent, multi-wavelength photometry ($U$-band to
8~$\mu$m) was measured using TFIT \citep{tfit}, implemented as
described by \citet{guo11} and Galametz et al. (2012, in prep.).
Photometric redshifts were computed using EAZY \citep{eazy} and
yielded errors of $\Delta z/(1+z)=3$\% and 6\% at $z>1.5$ in GOODS-S
and UDS, respectively. This dataset is partially described in
\citet{wuyts11b}; for full details, see Dahlen et al. (2012, in
prep.). Stellar masses were derived using FAST \citep{fast} and based
on a grid of \citet{bc03} models that assume a \citet{chabrier} IMF,
solar metallicity, exponentially declining star formation histories,
and the \citet{calzetti} dust extinction law.

SFRs were computed by combining IR and rest-frame UV (uncorrected for
extinction) luminosities (\citealt{ken98} and \citealt{bell05}) and
adopting a \citet{chabrier} IMF:
$SFR_{\mathrm{UV+IR}}=1.09\times10^{-10}(L_{\mathrm{IR}}+3.3L_{2800})$.
Total IR luminosities ($L_{\mathrm{IR}}$$\equiv L$[8-1000$\mu$m]) were
derived from \citet{ce01} templates fitting MIPS 24$\mu$m fluxes,
applying a {\it Herschel}-based re-calibration \citep{elbaz11}. For
galaxies undetected by MIPS below a 2$\sigma$ level (20$\mu$Jy) and
for galaxies detected in the X-rays, SFRs come from rest-frame UV
luminosities that are corrected for extinction as derived from SED
fits \citep{wuyts11a}. $L$$_{\mathrm{X}}\equiv L_{2-8\mathrm{kev}}$
were computed for the sources in the {\it Chandra} 4Ms image in
GOODS-S \citep{chandra4m} and the XMM 50--100ks survey in UDS
\citep{xmmuds}.  Due to the shallower detection limits in the IR and
X-ray surveys of UDS, the detection fractions are computed only on
GOODS-S data.

Circularized, effective (half-light) radii,
r$_{e}$$\equiv$$a\sqrt{(b/a)}$, and S\`ersic indices were measured
from {\it HST}/WFC3 $H$ images using GALFIT \citep{galfit} and PSFs
created and processed to replicate the conditions of the observed data
\citep[][and 2012 in prep]{vdw11a}.

\begin{figure*}
\centering
\includegraphics[width=18cm,angle=0.]{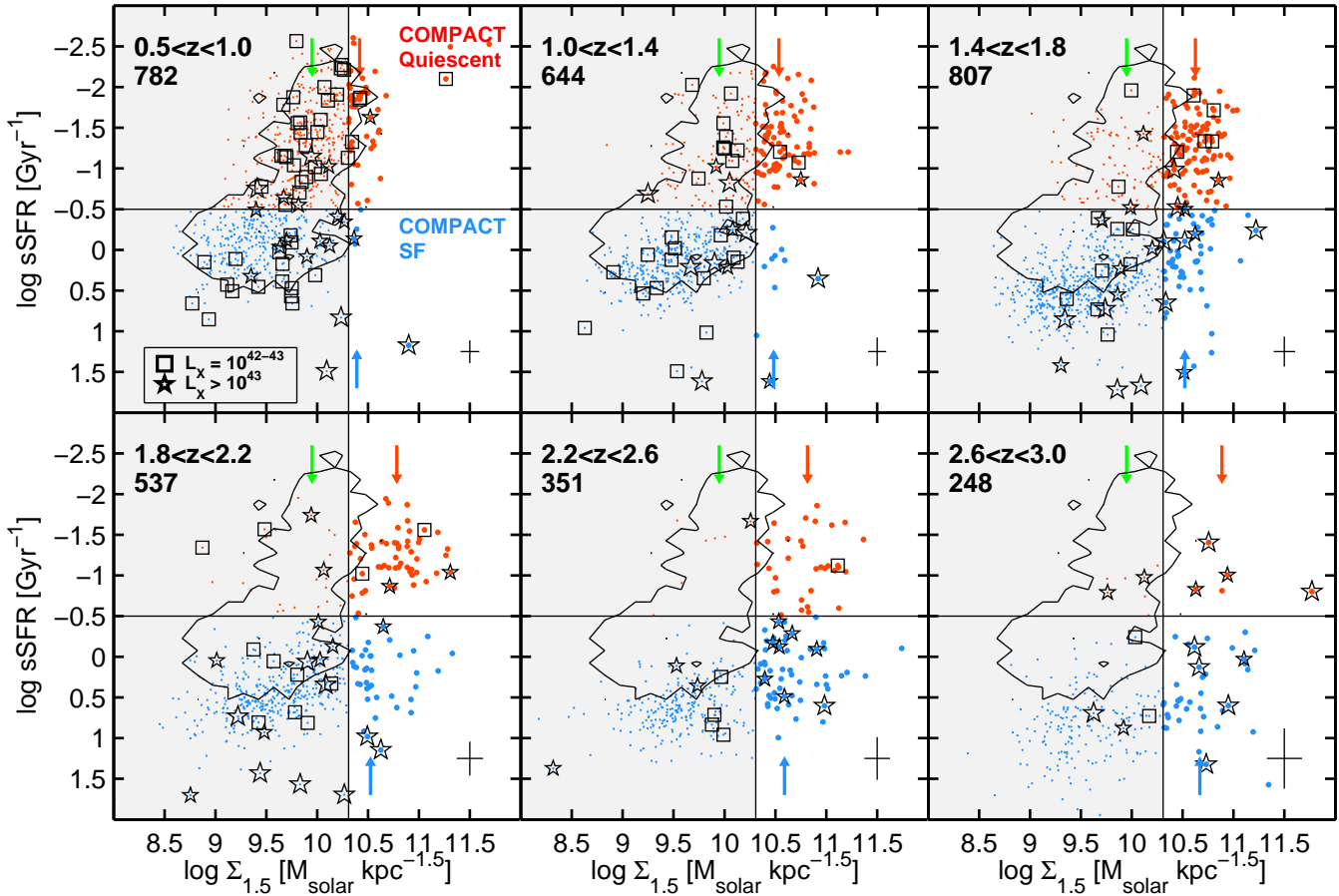}
\caption{\label{selection_diag} Evolution of the sSFR vs.
  $\Sigma_{1.5}$$\equiv$M/r$_{e}^{1.5}$ correlation at $0.5<z<3.0$ for
  galaxies above $M_{\star}>10^{10}M_{\odot}$. The redshift bins are
  chosen to probe similar comoving volumes. The solid lines and
  colored dots depict the selection thresholds for SFGs (blue) and QGs
  (red) and compact (white region) and non-compact (shaded region), as
  defined in Figure~\ref{ssfr_size}. The open markers depict sources
  detected in the X-rays at different luminosities: squares have
  $10^{42}<L_{\mathrm{X}}<10^{43}$~erg/s; small (large) stars have
  $L_{\mathrm{X}}>10^{43}$ ($10^{44}$; i.e, QSO) erg/s. Blue and red
  arrows indicate the median $\Sigma_{1.5}$ of cSFGs and cQGs,
  respectively. Green arrows approximate the local mass-size relation
  \citep{shen03}. The black contour shows the sSFR-$\Sigma_{1.5}$
  distribution for 90\% of the galaxies at $0.5<z<1.0$. The
  lower-right error bars for $\Sigma_{1.5}$ and the sSFR include
  uncertainties in: half-light radii, stellar masses, rest-frame
  luminosities (derived by perturbing photometric redshifts within the
  1$\sigma$ errors), and the average {\it rms} of the comparison
  between UV-corrected and (UV+IR)-based SFR estimates.}
\end{figure*}

We selected a sample of galaxies at $1.4<z<3.0$ with
$M_{\star}>10^{10}M_{\odot}$, above which our sample is $>$90\%
complete up to the highest redshifts (\citealt{wuyts11b};
\citealt{newman12}). Compact galaxies were based on $H$-band sizes;
quiescent galaxies (QGs) and star-forming galaxies (SFGs) were
separated by a specific SFR (sSFR) of 10$^{-0.5}$~Gyr$^{-1}$ (see
Figure~\ref{ssfr_size}). Although somewhat arbitrary, the value does
not strongly affect the results, since the sSFR bimodality is clearly
detected up to $z=3$.
  
Figure~\ref{ssfr_size} also shows a mass-size diagram for our sample.
In agreement with recent results, we find that SFGs and QGs follow
significantly different mass-size relations (\citealt{williams10};
\citealt{wuyts11b}). With this in mind, we select compact galaxies as
those following the observed trend in the mass-size relation for QGs
at $z>1.5$. The threshold is defined as
M/r$^{\alpha}_{\mathrm{e}}$$=$10.3M$_{\odot}$kpc$^{-\alpha}$, with
$\alpha=$1.5. The slope is roughly consistent with those given by
\citet{newman12} for QGs at similar redshifts
($\alpha^{-1}$=0.59-0.69). The zero-point is chosen to include the
majority of QGs with minimum contamination from SFGs.  For
$\alpha=$1.5, M/$r_{e}$$^{\alpha}$ (hereafter, $\Sigma_{1.5}$) lies
between the surface density, $\Sigma=$M/$r_{e}$$^{2}$, and M/$r_{e}$,
both of which follow strong correlations with color and SFR up to high
redshifts (e.g., \citealt{franx08}).

Figure~\ref{selection_diag} shows the evolution of sSFRs vs.
$\Sigma_{1.5}$ for massive galaxies from $z=3.0$ down to $z=0.5$. In
this diagram, our size-mass-SFR selection is completely orthogonal.
Although our analysis focuses on $z>1.4$, two panels at lower
redshifts are shown to illustrate the extrapolated evolutionary
trends. We find that the number of QGs (above the line) increases
rapidly since $z=3$, starting from very small number densities,
$n\sim10^{-5}$ Mpc$^{-3}$, at $z\sim2.8$. Among these, the number of
compact QGs (cQG; $\Sigma_{1.5}$$>$10.3) builds up first, and only at
$z<1.8$ we do start finding a sizable number of extended QGs. This
suggests that the bulk of these galaxies are assembled at late times
by both continuous migration (quenching) of non-compact SFGs
(bottom-left region) and size growth of cQGs. As a result of this
growth, the population of cQGs disappears by $z\sim1$. Simultaneously,
we identify a population of compact SFGs (cSFGs) whose number density
decreases steadily with time since $z=3.0$, being almost completely
absent at $z<1.4$. The number of cSFGs makes up less than 20\% of all
massive SFGs, but they present similar number densities as cQGs down
to $z\sim2$, suggesting an evolutionary link between the two
populations.

\section{Co-evolution of compact SFGs and QGs}

An evolutionary sequence where cSFGs are the progenitors of cQGs at
lower redshifts is supported by the fact that cSFGs first appear
before cQGs at high redshift ($z=2.6-3.0$) and then disappear before
them at low redshift ($z=1.0-1.4$), implying that evolution is from
blue through red rather than vice versa. Therefore, if we assume that
cSFGs would see their star formation quenched and fade at roughly
constant $\Sigma_{1.5}$, these could rapidly populate the compact
quiescent region in time scales of $\sim$500~Myr (approx. one of our
redshift intervals).

A more detailed analysis of the sizes of cSFGs and cQGs shows that,
indeed, both populations have median effective radii slightly smaller
than 1~kpc (similar to the findings in \citealt{vdw11a};
\citealt{szo11}). The median $\Sigma_{1.5}$ of cQGs (red arrows in
Figure~\ref{selection_diag}) decrease by 0.25~dex (i.e., increase
their radii by a factor of $\sim$2) from $z$$=$3.0 to $z$$=$1.4, in
agreement with previous results on the size evolution of QGs
(\citealt{cassata11}), whereas cSFGs present smaller values of
$\Sigma_{1.5}$ (blue arrows) by $\sim$0.2~dex, and a weaker evolution
with time. However, cSFGs migrating to the red sequence are expected
to slowly increase their masses with time, thus moving to higher
values of $\Sigma_{1.5}$ at lower redshifts.  Given their median sSFR,
the typical mass-doubling times for cSFGs range from 0.6 to
1~Gyr. This is enough to account for the observed difference in
$\Sigma_{1.5}$ between cSFGS and cQGs, provided that the newly formed
stars do not significantly increase the galaxy radii.

Both cQGs and cSFGs present similar surface brightness profiles, which
are best represented by large S\'ersic indices. The median values
range from $n=3-4$ for cQGs to $n=2.5-3.5$ for cSFGs. This means that
both populations are preferentially spheroid-like, in contrast with
non-compact, disk-like SFGs ($n\sim1$).  The median axis ratios of
cSFGs, b/a$\sim$0.65, are also consistent with spheroidal
morphologies.  However, cQGs present slightly smaller axis-ratios,
b/a$\sim$0.54, suggesting that some of these are small flattened disks
\citep{vdw11a}.  This feature might be explained if cQGs developed an
extended component surrounding the compact core, perhaps via minor
mergers (\citealt{naab09a}) or regrowth of a remnant disk that
survived the major merger (\citealt{governato09};
\citealt{hopkins09b}). These mechanisms will continue to growth these
galaxies is size, eventually depopulating the compact region.

Turning back to the possibility of cSFGs fading into cQGs, we find
that indeed cSFGs present suppressed sSFRs compared to the bulk of
SFGs. Although at $z>2$ the majority ($\sim$60--80\%) of cSFGs are
detected at 24$\mu$m, yielding SFR$=$100--200~M$_{\odot}$/yr$^{-1}$,
their median sSFRs are typically $\sim$0.3~dex lower than those for
non-compact SFGs at the same redshift. This suggests that the star
formation in the compact evolutionary stage already has started to
quench.

Simultaneously, we find an increasing fraction of cSFGs hosting X-ray
detected Active Galactic Nuclei (AGN) at $z>2$ (open markers in
Figure~\ref{selection_diag}). This result also supports the quenching
scenario, since AGN seems to be connected with quenching of the star
formation on time scales of a few hundred Myrs \citep{hopkins06}. In
particular, a high luminosity quasar phase
($L_{\mathrm{X}}$$>$10$^{44}$~erg~s$^{-1}$), associated with high
black-hole accretion rates, would be particularly efficient at
removing the available gas, thus stopping the star formation. Using
data from the deepest X-ray survey in GOODS-S, we find that cSFGs host
X-ray luminous AGNs 30x ($\sim$30\%) more frequently than non-compact
($<1$\%) (but massive) SFGs at $z\gtrsim2.2$. This implies that, at
these epochs, the majority of luminous
($L_{\mathrm{X}}$$>$10$^{43}$~erg/s) AGN are found preferentially in
compact hosts, as opposed to lower redshifts, where AGN are more
frequent in non-compact galaxies \citep{kocevski12}. Interestingly,
(the few) cQGs at $z\sim2.8$ also show a high fraction of X-ray
detections ($>70\%$), strengthening the idea that AGNs might play an
important role on the quenching of star formation.

Note that, although point source contamination from a luminous AGN
could bias the structural parameters towards compact morphologies
\citep{pierce10}, X-ray detected galaxies do not deviate from the
general trend followed by non-Xray detected cSFGs in
Figure~\ref{selection_diag}, with the exception of a few extreme
outliers. These AGNs and also those with strong SED contamination were
excluded from our sample prior to the analysis described above.

\begin{figure}
\centering
\includegraphics[width=8.5cm,angle=0.]{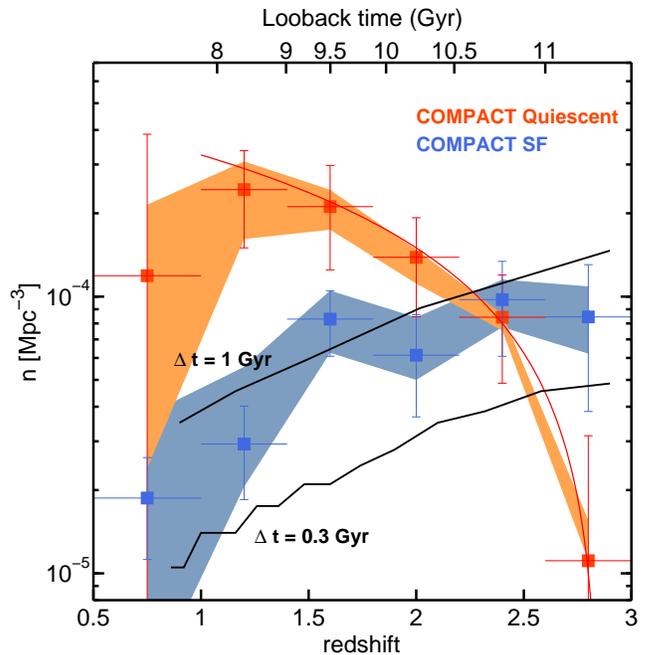}
\caption{\label{density} Number density evolution of massive,
  $M_{\star}>10^{10}M_{\odot}$, cQGs (red) and cSFGs (blue) versus
  redshift. The solid red line is the best fit to the number density
  of cQGs. Solid black lines depict the evolution of the number of
  cSFGs required to match the observed increasing density of cQGs,
  assuming that the former have lifetimes of $\Delta
  t_{burst}=0.3-1$~Gyr. The error bars were computed by bootstrapping
  the sSFR and $\Sigma_{1.5}$ uncertainties along with terms for small
  number statistics and field-to-field differences.  The shaded
  regions encompass the observed number densities when the selection
  thresholds in sSFR and $\Sigma_{1.5}$ are modified by $\pm$0.2~dex.}
\end{figure}

\section{Number density of compact galaxies}

Figure~\ref{density} shows the number density evolution of massive
cQGs and cSFGs. The best fit (red line) to the increasing number
density of QGs can be parametrized as $n=$a$+$b(1+z), with
a$=$1.75$\times$10$^{-4}$ and b$=$-6.75$\times$10$^{-4}$. Assuming
that all cQGs at a given redshift $z'$ come from cSFGs, we can
estimate how many of the latter we should observe at $z>z'$.  To do
so, we propose a simple evolutionary model that assigns to all SFGs an
arbitrary lifetime for their current burst of star formation, $\Delta
t_{burst}$, after which they will become quiescent. The number of cQGs
at a given time would be:

\begin{equation}
n_{\mathrm{QG}}(t+\Delta t_{burst})=n_{\mathrm{QG}}(t)+n_{\mathrm{SFG}}(t) 
\end{equation}

We explore $\Delta t_{burst}$ values from 0.3~Gyr to 1.0~Gyr, similar to
the typical e-folding times expected for SFGs at these redshifts
\citep{wuyts11a}.  The observed number density of cSFGs is broadly
consistent with the model prediction for a median value of
$\Delta t_{burst}$$\sim$800~Myr. 

This simple model assumes, that at every step, $\Delta t_{burst}$,
enough cSFGs are being formed by some mechanism(s), restoring the ones
that turned into cQGs.  Plausible mechanisms to reduce the size of
massive (larger) SFGs are gas-rich dissipational processes, such as
mergers or dynamical instabilities. These can produce compact,
star-bursting remnants that would likely quench in a short period of
time (\citealt{hopkins06}; \citealt{dekel09}). Without relying on
mergers being the main or sole driver of this transformation, we can
make a quantitative estimate of the number of cSFGs assembled by this
mechanism by using typical numbers for major mergers at these
redshifts. Considering pair fractions of roughly 10\%
\citep{williams11}, merger time scales of 1~Gyr (\citealt{lotz11}) and
a density of massive galaxies of $\lesssim10^{-3}$~Mpc$^{-3}$
(\citealt{pg08}), we obtain an assembly rate for new cSFGs via mergers
of $\Delta n_\mathrm{cSFG}$$\sim$10$^{-4}$~Gyr$^{-1}$, which is
roughly consistent with the observed densities for the predicted
$\Delta t_{burst}$.

In this model, the significant decrement on the number cSFGs by
$z<1.4$ implies that the formation mechanism(s) become quickly
inefficient at lower redshifts, thereby truncating the formation of
new cSFGs and thus cQGs.  This gradual decline of the efficiency of
the dissipational processes may follow the decline in the amount of
available gas in dark matter haloes (e.g.,
\citealt{croton09}). Detailed comparisons to cosmological models will
allow tests of this hypothesis and provide a more rigorous modeling of
the number density evolution (Porter et al., in prep). Also, a follow
up survey of cSFGs at $z\gtrsim$3 will place better constraints on the
formation scenario for these galaxies (C. Williams et al., in prep).

\section{Summary and discussion}

\begin{figure}
\centering \includegraphics[width=8.5cm,angle=0.]{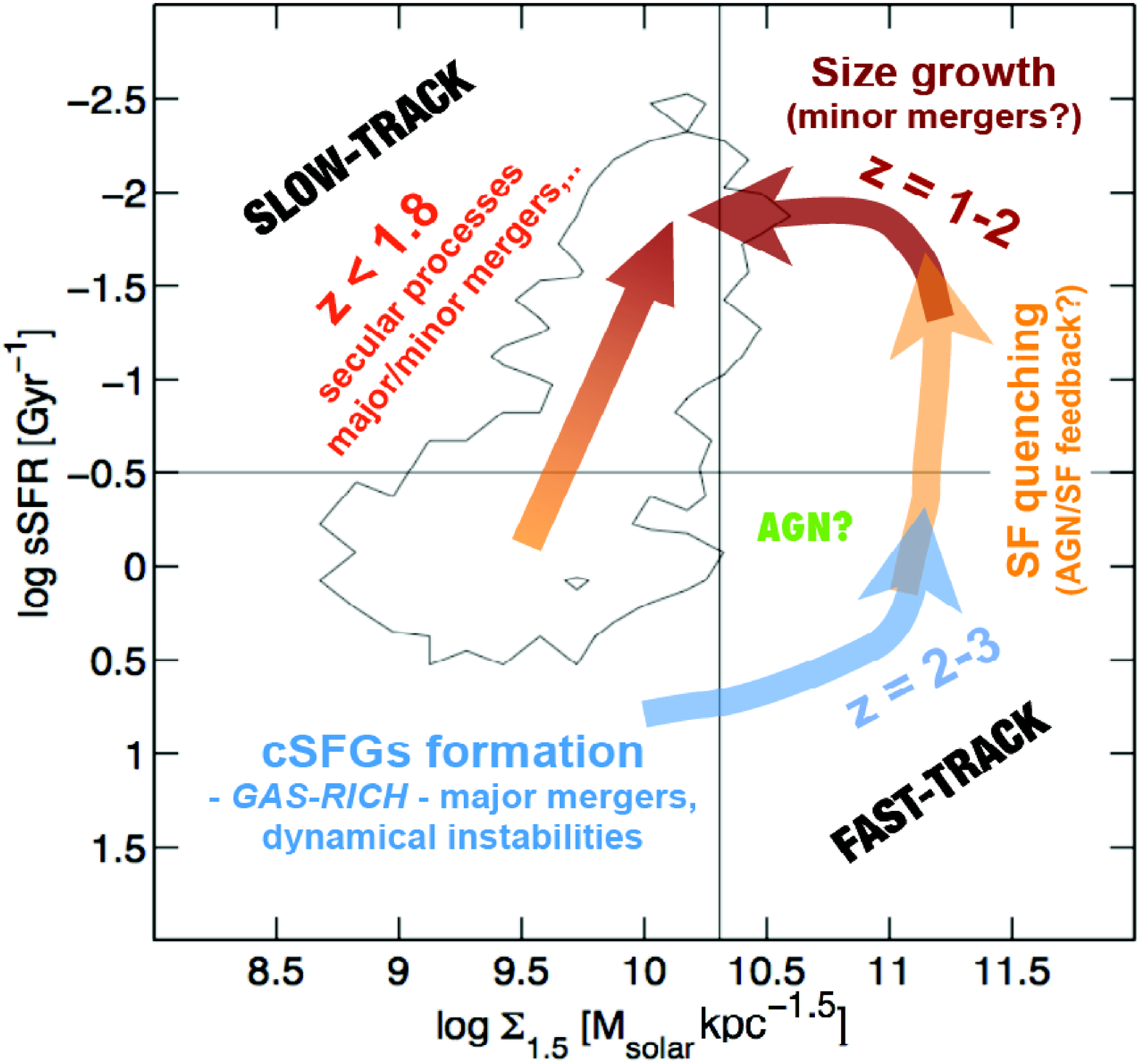}
\caption{\label{cartoon} Schematic view of a two path (fast/slow
  track) formation scenario for QGs. On the fast track, a small
  fraction of the massive SFGs at $z=2-3$ evolve (e.g., through
  gas-rich dissipational processes) to a compact star-bursting
  remnant.  Then, the star formation is quenched in $\sim$800~Myr (
  perhaps by AGN and/or supernovae feedback), and galaxies fade into
  cQGs. Once in the red sequence, cQGs grow envelopes, over longer
  time scales, de-populating the compact region by
  z$\sim$1. Simultaneously, at $z\lesssim2$, other (slower) mechanisms
  have already started to populate the red-sequence with normal-sized,
  non-compact QGs (formed by, e.g., secular processes, halo quenching,
  or gas-poor mergers)}
\end{figure}

Using the deepest data spanning from the X-ray-to-MIR, along with high
resolution imaging from CANDELS in GOODS-S and UDS, we analyze stellar
masses, SFRs and sizes of a sample of massive
($M_{\star}>10^{10}M_{\odot}$) galaxies at $z=1.4-3.0$ to identify a
population of cSFGs with similar structural properties as cQGs at
$z\gtrsim2$.  The cSFG population is already in place at $z\sim3$, but
it completely disappears by $z<1.4$. A corresponding increase in the
number of cQGs during the same time period suggests an evolutionary
link between them.

A simple duty-cycle argument, involving quenching of the star
formation activity on time scales of $\Delta t=0.3-1$~Gyr, is able to
broadly reproduce the evolution of the density of new QGs formed since
$z=3$ in terms of fading cSFGs. Under this assumption, we also need to
invoke a replenishment mechanism to form new cSFG via gas-rich
dissipational processes (major mergers or dynamical instabilities),
that then become quickly inefficient at $z\lesssim1.5$, as the amount
of available gas in the halo decreases with time (e.g.,
\citealt{croton09}).

During the transformation processes, the compact phase is probably
associated with: enhanced (probably nuclear and dusty) star formation,
the presence of an AGN, and sometimes a short-lived quasar, followed
by a decline of the star formation in $\sim$1~Gyr
\citep{hopkins06}. All these phenomena fit with the observed
properties of cSFGs presented in this letter. cSFGs present no visible
traces of mergers, but they do show lower sSFRs than the bulk of
massive SFGs, presumably being at different stages of the starburst to
passive evolution. Simultaneously, $\sim$30\% of them host luminous
($L_{X}>10^{43}$erg~s$^{-1}$) X-ray detected AGNs at $z>2$, suggesting
that these might be playing a role in the quenching of star formation.

Our observations connect two recent results at $z\sim2$: a population
of compact ($n\gtrsim2$) galaxies with enhanced star formation
activity \citep{wuyts11b} and an increasing fraction of small,
post-starburst galaxies recently arrived on the red sequence
\citep{whitaker12}. The emerging picture suggests that the formation
of QGs follows two evolutionary tracks, each one dominating at
different epochs (as illustrated in Figure~\ref{cartoon}). At
$z\gtrsim2$ the formation of QGs proceeds on a fast track
(right-region): from $z=3.0-2.0$, the number of cQGs builds up rapidly
upon quenching of cSFGs at roughly constant $\Sigma_{1.5}$. Merely
2~Gyrs later ($z\sim1$), cQGs almost completely disappear due to: 1)
size growth as a result of minor mergers satellite accretion
(\citealt{naab09a}; \citealt{newman12}) or the re-growth of a remnant
disk (\citealt{hopkins09b}) which causes them to leave the compact
region; 2) a decrement in the efficiency of the formation mechanisms
for new cSFGs, and therefore new cQGs.  By $z\lesssim2$, other
(probably slower) mechanisms start to populate the red sequence with
larger, non-compact, QGs without passing through a compact state. This
slow track (left region) is likely associated with the fading of
normal disk galaxies to become S0s, due to halo quenching
\citep{haloquench} or morphological quenching
\citep{martig09}. Alternatively, some of these QGs could also be the
result late and less-gas-rich mergers.

\section*{Acknowledgments}
Support for Program number HST-GO-12060 was provided by NASA through a
grant from the Space Telescope Science Institute, which is operated by
the Association of Universities for Research in Astronomy,
Incorporated, under NASA contract NAS5-26555. PGP-G acknowledges
support from grant AYA2009-07723-E.

\bibliographystyle{aa}

\begin{thebibliography}{50}
\expandafter\ifx\csname natexlab\endcsname\relax\def\natexlab#1{#1}\fi

\bibitem[{{Bell} {et~al.}(2005){Bell}, {Papovich}, {Wolf}, {Le Floc'h},
  {Caldwell}, {Barden}, {Egami}, {McIntosh}, {Meisenheimer},
  {P{\'e}rez-Gonz{\'a}lez}, {Rieke}, {Rieke}, {Rigby}, \& {Rix}}]{bell05}
{Bell}, E.~F., {Papovich}, C., {Wolf}, C., {et~al.} 2005, \apj, 625, 23

\bibitem[{{Bell} {et~al.}(2011){Bell}, {van der Wel}, {Papovich}, {Kocevski},
  {Lotz}, {McIntosh}, {Kartaltepe}, {Faber}, {Ferguson}, {Koekemoer}, {Grogin},
  {Wuyts}, {Cheung}, {Conselice}, {Dekel}, {Dunlop}, {Giavalisco},
  {Herrington}, {Koo}, {McGrath}, {de Mello}, {Rix}, {Robaina}, \&
  {Williams}}]{bell11}
{Bell}, E.~F., {van der Wel}, A., {Papovich}, C., {et~al.} 2011, ArXiv e-prints

\bibitem[{{Birnboim} \& {Dekel}(2003)}]{haloquench}
{Birnboim}, Y. \& {Dekel}, A. 2003, \mnras, 345, 349

\bibitem[{{Brammer} {et~al.}(2008){Brammer}, {van Dokkum}, \& {Coppi}}]{eazy}
{Brammer}, G.~B., {van Dokkum}, P.~G., \& {Coppi}, P. 2008, \apj, 686, 1503

\bibitem[{{Brammer} {et~al.}(2011){Brammer}, {Whitaker}, {van Dokkum},
  {Marchesini}, {Franx}, {Kriek}, {Labb{\'e}}, {Lee}, {Muzzin}, {Quadri},
  {Rudnick}, \& {Williams}}]{brammer11}
{Brammer}, G.~B., {Whitaker}, K.~E., {van Dokkum}, P.~G., {et~al.} 2011, \apj,
  739, 24

\bibitem[{{Bruzual} \& {Charlot}(2003)}]{bc03}
{Bruzual}, G. \& {Charlot}, S. 2003, \mnras, 344, 1000

\bibitem[{{Calzetti} {et~al.}(2000){Calzetti}, {Armus}, {Bohlin}, {Kinney},
  {Koornneef}, \& {Storchi-Bergmann}}]{calzetti}
{Calzetti}, D., {Armus}, L., {Bohlin}, R.~C., {et~al.} 2000, \apj, 533, 682

\bibitem[{{Cassata} {et~al.}(2011){Cassata}, {Giavalisco}, {Guo}, {Renzini},
  {Ferguson}, {Koekemoer}, {Salimbeni}, {Scarlata}, {Grogin}, {Conselice},
  {Dahlen}, {Lotz}, {Dickinson}, \& {Lin}}]{cassata11}
{Cassata}, P., {Giavalisco}, M., {Guo}, Y., {et~al.} 2011, \apj, 743, 96

\bibitem[{{Cava} {et~al.}(2010){Cava}, {Rodighiero}, {P{\'e}rez-Fournon},
  {Buitrago}, {Trujillo}, {Altieri}, {Amblard}, {Auld}, {Bock}, {Brisbin},
  {Burgarella}, {Castro-Rodr{\'{\i}}guez}, {Chanial}, {Cirasuolo}, {Clements},
  {Conselice}, {Cooray}, {Eales}, {Elbaz}, {Ferrero}, {Franceschini}, {Glenn},
  {Solares}, {Griffin}, {Ibar}, {Ivison}, {Marchetti}, {Morrison}, {Mortier},
  {Oliver}, {Page}, {Papageorgiou}, {Pearson}, {Pohlen}, {Rawlings}, {Raymond},
  {Rigopoulou}, {Roseboom}, {Rowan-Robinson}, {Scott}, {Seymour}, {Smith},
  {Symeonidis}, {Tugwell}, {Vaccari}, {Valtchanov}, {Vieira}, {Vigroux},
  {Wang}, \& {Wright}}]{cava03}
{Cava}, A., {Rodighiero}, G., {P{\'e}rez-Fournon}, I., {et~al.} 2010, \mnras,
  409, L19

\bibitem[{{Chabrier}(2003)}]{chabrier}
{Chabrier}, G. 2003, \pasp, 115, 763

\bibitem[{{Chary} \& {Elbaz}(2001)}]{ce01}
{Chary}, R. \& {Elbaz}, D. 2001, \apj, 556, 562

\bibitem[{{Croton}(2009)}]{croton09}
{Croton}, D.~J. 2009, \mnras, 394, 1109

\bibitem[{{Dekel} {et~al.}(2009){Dekel}, {Sari}, \& {Ceverino}}]{dekel09}
{Dekel}, A., {Sari}, R., \& {Ceverino}, D. 2009, \apj, 703, 785

\bibitem[{{Elbaz} {et~al.}(2011){Elbaz}, {Dickinson}, {Hwang},
  {D{\'{\i}}az-Santos}, {Magdis}, {Magnelli}, {Le Borgne}, {Galliano},
  {Pannella}, {Chanial}, {Armus}, {Charmandaris}, {Daddi}, {Aussel}, {Popesso},
  {Kartaltepe}, {Altieri}, {Valtchanov}, {Coia}, {Dannerbauer}, {Dasyra},
  {Leiton}, {Mazzarella}, {Alexander}, {Buat}, {Burgarella}, {Chary}, {Gilli},
  {Ivison}, {Juneau}, {Le Floc'h}, {Lutz}, {Morrison}, {Mullaney}, {Murphy},
  {Pope}, {Scott}, {Brodwin}, {Calzetti}, {Cesarsky}, {Charlot}, {Dole},
  {Eisenhardt}, {Ferguson}, {F{\"o}rster Schreiber}, {Frayer}, {Giavalisco},
  {Huynh}, {Koekemoer}, {Papovich}, {Reddy}, {Surace}, {Teplitz}, {Yun}, \&
  {Wilson}}]{elbaz11}
{Elbaz}, D., {Dickinson}, M., {Hwang}, H.~S., {et~al.} 2011, \aap, 533, A119

\bibitem[{{Franx} {et~al.}(2008){Franx}, {van Dokkum}, {Schreiber}, {Wuyts},
  {Labb{\'e}}, \& {Toft}}]{franx08}
{Franx}, M., {van Dokkum}, P.~G., {Schreiber}, N.~M.~F., {et~al.} 2008, \apj,
  688, 770

\bibitem[{{Giavalisco} {et~al.}(2004){Giavalisco}, {Ferguson}, {Koekemoer},
  {Dickinson}, {Alexander}, {Bauer}, {Bergeron}, {Biagetti}, {Brandt},
  {Casertano}, {Cesarsky}, {Chatzichristou}, {Conselice}, {Cristiani}, {Da
  Costa}, {Dahlen}, {de Mello}, {Eisenhardt}, {Erben}, {Fall}, {Fassnacht},
  {Fosbury}, {Fruchter}, {Gardner}, {Grogin}, {Hook}, {Hornschemeier}, {Idzi},
  {Jogee}, {Kretchmer}, {Laidler}, {Lee}, {Livio}, {Lucas}, {Madau},
  {Mobasher}, {Moustakas}, {Nonino}, {Padovani}, {Papovich}, {Park},
  {Ravindranath}, {Renzini}, {Richardson}, {Riess}, {Rosati}, {Schirmer},
  {Schreier}, {Somerville}, {Spinrad}, {Stern}, {Stiavelli}, {Strolger},
  {Urry}, {Vandame}, {Williams}, \& {Wolf}}]{goods}
{Giavalisco}, M., {Ferguson}, H.~C., {Koekemoer}, A.~M., {et~al.} 2004, \apjl,
  600, L93

\bibitem[{{Governato} {et~al.}(2009){Governato}, {Brook}, {Brooks}, {Mayer},
  {Willman}, {Jonsson}, {Stilp}, {Pope}, {Christensen}, {Wadsley}, \&
  {Quinn}}]{governato09}
{Governato}, F., {Brook}, C.~B., {Brooks}, A.~M., {et~al.} 2009, \mnras, 398,
  312

\bibitem[{{Grogin} {et~al.}(2011){Grogin}, {Kocevski}, {Faber}, {Ferguson},
  {Koekemoer}, {Riess}, {Acquaviva}, {Alexander}, {Almaini}, {Ashby}, {Barden},
  {Bell}, {Bournaud}, {Brown}, {Caputi}, {Casertano}, {Cassata}, {Castellano},
  {Challis}, {Chary}, {Cheung}, {Cirasuolo}, {Conselice}, {Roshan Cooray},
  {Croton}, {Daddi}, {Dahlen}, {Dav{\'e}}, {de Mello}, {Dekel}, {Dickinson},
  {Dolch}, {Donley}, {Dunlop}, {Dutton}, {Elbaz}, {Fazio}, {Filippenko},
  {Finkelstein}, {Fontana}, {Gardner}, {Garnavich}, {Gawiser}, {Giavalisco},
  {Grazian}, {Guo}, {Hathi}, {H{\"a}ussler}, {Hopkins}, {Huang}, {Huang},
  {Jha}, {Kartaltepe}, {Kirshner}, {Koo}, {Lai}, {Lee}, {Li}, {Lotz}, {Lucas},
  {Madau}, {McCarthy}, {McGrath}, {McIntosh}, {McLure}, {Mobasher},
  {Moustakas}, {Mozena}, {Nandra}, {Newman}, {Niemi}, {Noeske}, {Papovich},
  {Pentericci}, {Pope}, {Primack}, {Rajan}, {Ravindranath}, {Reddy}, {Renzini},
  {Rix}, {Robaina}, {Rodney}, {Rosario}, {Rosati}, {Salimbeni}, {Scarlata},
  {Siana}, {Simard}, {Smidt}, {Somerville}, {Spinrad}, {Straughn}, {Strolger},
  {Telford}, {Teplitz}, {Trump}, {van der Wel}, {Villforth}, {Wechsler},
  {Weiner}, {Wiklind}, {Wild}, {Wilson}, {Wuyts}, {Yan}, \& {Yun}}]{candelsgro}
{Grogin}, N.~A., {Kocevski}, D.~D., {Faber}, S.~M., {et~al.} 2011, \apjs, 197,
  35

\bibitem[{{Guo} {et~al.}(2011){Guo}, {Giavalisco}, {Cassata}, {Ferguson},
  {Dickinson}, {Renzini}, {Koekemoer}, {Grogin}, {Papovich}, {Tundo},
  {Fontana}, {Lotz}, \& {Salimbeni}}]{guo11}
{Guo}, Y., {Giavalisco}, M., {Cassata}, P., {et~al.} 2011, \apj, 735, 18

\bibitem[{{Hopkins} {et~al.}(2006){Hopkins}, {Hernquist}, {Cox}, {Di Matteo},
  {Robertson}, \& {Springel}}]{hopkins06}
{Hopkins}, P.~F., {Hernquist}, L., {Cox}, T.~J., {et~al.} 2006, \apjs, 163, 1

\bibitem[{{Hopkins} {et~al.}(2009){Hopkins}, {Hernquist}, {Cox}, {Keres}, \&
  {Wuyts}}]{hopkins09b}
{Hopkins}, P.~F., {Hernquist}, L., {Cox}, T.~J., {Keres}, D., \& {Wuyts}, S.
  2009, \apj, 691, 1424

\bibitem[{{Ilbert} {et~al.}(2010){Ilbert}, {Salvato}, {Le Floc'h}, {Aussel},
  {Capak}, {McCracken}, {Mobasher}, {Kartaltepe}, {Scoville}, {Sanders},
  {Arnouts}, {Bundy}, {Cassata}, {Kneib}, {Koekemoer}, {Le F{\`e}vre}, {Lilly},
  {Surace}, {Taniguchi}, {Tasca}, {Thompson}, {Tresse}, {Zamojski}, {Zamorani},
  \& {Zucca}}]{2010ApJ...709..644I}
{Ilbert}, O., {Salvato}, M., {Le Floc'h}, E., {et~al.} 2010, \apj, 709, 644

\bibitem[{{Kauffmann} {et~al.}(2003){Kauffmann}, {Heckman}, {White}, {Charlot},
  {Tremonti}, {Peng}, {Seibert}, {Brinkmann}, {Nichol}, {SubbaRao}, \&
  {York}}]{kauffman03}
{Kauffmann}, G., {Heckman}, T.~M., {White}, S.~D.~M., {et~al.} 2003, \mnras,
  341, 54

\bibitem[{{Kennicutt}(1998)}]{ken98}
{Kennicutt}, Jr., R.~C. 1998, \araa, 36, 189

\bibitem[{{Kocevski} {et~al.}(2012){Kocevski}, {Faber}, {Mozena}, {Koekemoer},
  {Nandra}, {Rangel}, {Laird}, {Brusa}, {Wuyts}, {Trump}, {Koo}, {Somerville},
  {Bell}, {Lotz}, {Alexander}, {Bournaud}, {Conselice}, {Dahlen}, {Dekel},
  {Donley}, {Dunlop}, {Finoguenov}, {Georgakakis}, {Giavalisco}, {Guo},
  {Grogin}, {Hathi}, {Juneau}, {Kartaltepe}, {Lucas}, {McGrath}, {McIntosh},
  {Mobasher}, {Robaina}, {Rosario}, {Straughn}, {van der Wel}, \&
  {Villforth}}]{kocevski12}
{Kocevski}, D.~D., {Faber}, S.~M., {Mozena}, M., {et~al.} 2012, \apj, 744, 148

\bibitem[{{Koekemoer} {et~al.}(2011){Koekemoer}, {Faber}, {Ferguson}, {Grogin},
  {Kocevski}, {Koo}, {Lai}, {Lotz}, {Lucas}, {McGrath}, {Ogaz}, {Rajan},
  {Riess}, {Rodney}, {Strolger}, {Casertano}, {Castellano}, {Dahlen},
  {Dickinson}, {Dolch}, {Fontana}, {Giavalisco}, {Grazian}, {Guo}, {Hathi},
  {Huang}, {van der Wel}, {Yan}, {Acquaviva}, {Alexander}, {Almaini}, {Ashby},
  {Barden}, {Bell}, {Bournaud}, {Brown}, {Caputi}, {Cassata}, {Challis},
  {Chary}, {Cheung}, {Cirasuolo}, {Conselice}, {Roshan Cooray}, {Croton},
  {Daddi}, {Dav{\'e}}, {de Mello}, {de Ravel}, {Dekel}, {Donley}, {Dunlop},
  {Dutton}, {Elbaz}, {Fazio}, {Filippenko}, {Finkelstein}, {Frazer}, {Gardner},
  {Garnavich}, {Gawiser}, {Gruetzbauch}, {Hartley}, {H{\"a}ussler},
  {Herrington}, {Hopkins}, {Huang}, {Jha}, {Johnson}, {Kartaltepe},
  {Khostovan}, {Kirshner}, {Lani}, {Lee}, {Li}, {Madau}, {McCarthy},
  {McIntosh}, {McLure}, {McPartland}, {Mobasher}, {Moreira}, {Mortlock},
  {Moustakas}, {Mozena}, {Nandra}, {Newman}, {Nielsen}, {Niemi}, {Noeske},
  {Papovich}, {Pentericci}, {Pope}, {Primack}, {Ravindranath}, {Reddy},
  {Renzini}, {Rix}, {Robaina}, {Rosario}, {Rosati}, {Salimbeni}, {Scarlata},
  {Siana}, {Simard}, {Smidt}, {Snyder}, {Somerville}, {Spinrad}, {Straughn},
  {Telford}, {Teplitz}, {Trump}, {Vargas}, {Villforth}, {Wagner}, {Wandro},
  {Wechsler}, {Weiner}, {Wiklind}, {Wild}, {Wilson}, {Wuyts}, \&
  {Yun}}]{candelskoe}
{Koekemoer}, A.~M., {Faber}, S.~M., {Ferguson}, H.~C., {et~al.} 2011, \apjs,
  197, 36

\bibitem[{{Kriek} {et~al.}(2009{\natexlab{a}}){Kriek}, {van Dokkum}, {Franx},
  {Illingworth}, \& {Magee}}]{kriek09}
{Kriek}, M., {van Dokkum}, P.~G., {Franx}, M., {Illingworth}, G.~D., \&
  {Magee}, D.~K. 2009{\natexlab{a}}, \apjl, 705, L71

\bibitem[{{Kriek} {et~al.}(2009{\natexlab{b}}){Kriek}, {van Dokkum},
  {Labb{\'e}}, {Franx}, {Illingworth}, {Marchesini}, \& {Quadri}}]{fast}
{Kriek}, M., {van Dokkum}, P.~G., {Labb{\'e}}, I., {et~al.} 2009{\natexlab{b}},
  \apj, 700, 221

\bibitem[{{Laidler} {et~al.}(2006){Laidler}, {Grogin}, {Clubb}, {Ferguson},
  {Papovich}, {Dickinson}, {Idzi}, {MacDonald}, {Ouchi}, \& {Mobasher}}]{tfit}
{Laidler}, V.~G., {Grogin}, N., {Clubb}, K., {et~al.} 2006, in Astronomical
  Society of the Pacific Conference Series, Vol. 351, Astronomical Data
  Analysis Software and Systems XV, ed. C.~{Gabriel}, C.~{Arviset}, D.~{Ponz},
  \& S.~{Enrique}, 228

\bibitem[{{Lotz} {et~al.}(2011){Lotz}, {Jonsson}, {Cox}, {Croton}, {Primack},
  {Somerville}, \& {Stewart}}]{lotz11}
{Lotz}, J.~M., {Jonsson}, P., {Cox}, T.~J., {et~al.} 2011, \apj, 742, 103

\bibitem[{{Martig} {et~al.}(2009){Martig}, {Bournaud}, {Teyssier}, \&
  {Dekel}}]{martig09}
{Martig}, M., {Bournaud}, F., {Teyssier}, R., \& {Dekel}, A. 2009, \apj, 707,
  250

\bibitem[{{Naab} {et~al.}(2007){Naab}, {Johansson}, {Ostriker}, \&
  {Efstathiou}}]{naab07}
{Naab}, T., {Johansson}, P.~H., {Ostriker}, J.~P., \& {Efstathiou}, G. 2007,
  \apj, 658, 710

\bibitem[{{Naab} \& {Ostriker}(2009)}]{naab09a}
{Naab}, T. \& {Ostriker}, J.~P. 2009, \apj, 690, 1452

\bibitem[{{Newman} {et~al.}(2012){Newman}, {Ellis}, {Bundy}, \&
  {Treu}}]{newman12}
{Newman}, A.~B., {Ellis}, R.~S., {Bundy}, K., \& {Treu}, T. 2012, \apj, 746,
  162

\bibitem[{{Peng} {et~al.}(2002){Peng}, {Ho}, {Impey}, \& {Rix}}]{galfit}
{Peng}, C.~Y., {Ho}, L.~C., {Impey}, C.~D., \& {Rix}, H.-W. 2002, \aj, 124, 266

\bibitem[{{P{\'e}rez-Gonz{\'a}lez} {et~al.}(2008){P{\'e}rez-Gonz{\'a}lez},
  {Rieke}, {Villar}, {Barro}, {Blaylock}, {Egami}, {Gallego}, {Gil de Paz},
  {Pascual}, {Zamorano}, \& {Donley}}]{pg08}
{P{\'e}rez-Gonz{\'a}lez}, P.~G., {Rieke}, G.~H., {Villar}, V., {et~al.} 2008,
  \apj, 675, 234

\bibitem[{{Pierce} {et~al.}(2010){Pierce}, {Lotz}, {Primack}, {Rosario},
  {Griffith}, {Conselice}, {Faber}, {Koo}, {Coil}, {Salim}, {Koekemoer},
  {Laird}, {Ivison}, \& {Yan}}]{pierce10}
{Pierce}, C.~M., {Lotz}, J.~M., {Primack}, J.~R., {et~al.} 2010, \mnras, 405,
  718

\bibitem[{{Shen} {et~al.}(2003){Shen}, {Mo}, {White}, {Blanton}, {Kauffmann},
  {Voges}, {Brinkmann}, \& {Csabai}}]{shen03}
{Shen}, S., {Mo}, H.~J., {White}, S.~D.~M., {et~al.} 2003, \mnras, 343, 978

\bibitem[{{Szomoru} {et~al.}(2011){Szomoru}, {Franx}, {Bouwens}, {van Dokkum},
  {Labb{\'e}}, {Illingworth}, \& {Trenti}}]{szo11}
{Szomoru}, D., {Franx}, M., {Bouwens}, R.~J., {et~al.} 2011, \apjl, 735, L22

\bibitem[{{Tacconi} {et~al.}(2010){Tacconi}, {Genzel}, {Neri}, {Cox}, {Cooper},
  {Shapiro}, {Bolatto}, {Bouch{\'e}}, {Bournaud}, {Burkert}, {Combes},
  {Comerford}, {Davis}, {Schreiber}, {Garcia-Burillo}, {Gracia-Carpio}, {Lutz},
  {Naab}, {Omont}, {Shapley}, {Sternberg}, \& {Weiner}}]{tacconi10}
{Tacconi}, L.~J., {Genzel}, R., {Neri}, R., {et~al.} 2010, \nat, 463, 781

\bibitem[{{Trujillo} {et~al.}(2007){Trujillo}, {Conselice}, {Bundy}, {Cooper},
  {Eisenhardt}, \& {Ellis}}]{2007MNRAS.382..109T}
{Trujillo}, I., {Conselice}, C.~J., {Bundy}, K., {et~al.} 2007, \mnras, 382,
  109

\bibitem[{{Ueda} {et~al.}(2008){Ueda}, {Watson}, {Stewart}, {Akiyama},
  {Schwope}, {Lamer}, {Ebrero}, {Carrera}, {Sekiguchi}, {Yamada}, {Simpson},
  {Hasinger}, \& {Mateos}}]{xmmuds}
{Ueda}, Y., {Watson}, M.~G., {Stewart}, I.~M., {et~al.} 2008, \apjs, 179, 124

\bibitem[{{van der Wel} {et~al.}(2011){van der Wel}, {Rix}, {Wuyts}, {McGrath},
  {Koekemoer}, {Bell}, {Holden}, {Robaina}, \& {McIntosh}}]{vdw11a}
{van der Wel}, A., {Rix}, H.-W., {Wuyts}, S., {et~al.} 2011, \apj, 730, 38

\bibitem[{{Whitaker} {et~al.}(2012){Whitaker}, {Kriek}, {van Dokkum},
  {Bezanson}, {Brammer}, {Franx}, \& {Labb{\'e}}}]{whitaker12}
{Whitaker}, K.~E., {Kriek}, M., {van Dokkum}, P.~G., {et~al.} 2012, \apj, 745,
  179

\bibitem[{{Williams} {et~al.}(2011){Williams}, {Quadri}, \&
  {Franx}}]{williams11}
{Williams}, R.~J., {Quadri}, R.~F., \& {Franx}, M. 2011, \apjl, 738, L25

\bibitem[{{Williams} {et~al.}(2010){Williams}, {Quadri}, {Franx}, {van Dokkum},
  {Toft}, {Kriek}, \& {Labb{\'e}}}]{williams10}
{Williams}, R.~J., {Quadri}, R.~F., {Franx}, M., {et~al.} 2010, \apj, 713, 738

\bibitem[{{Wuyts} {et~al.}(2010){Wuyts}, {Cox}, {Hayward}, {Franx},
  {Hernquist}, {Hopkins}, {Jonsson}, \& {van Dokkum}}]{wuyts10}
{Wuyts}, S., {Cox}, T.~J., {Hayward}, C.~C., {et~al.} 2010, \apj, 722, 1666

\bibitem[{{Wuyts} {et~al.}(2011{\natexlab{a}}){Wuyts}, {F{\"o}rster Schreiber},
  {Lutz}, {Nordon}, {Berta}, {Altieri}, {Andreani}, {Aussel}, {Bongiovanni},
  {Cepa}, {Cimatti}, {Daddi}, {Elbaz}, {Genzel}, {Koekemoer}, {Magnelli},
  {Maiolino}, {McGrath}, {P{\'e}rez Garc{\'{\i}}a}, {Poglitsch}, {Popesso},
  {Pozzi}, {Sanchez-Portal}, {Sturm}, {Tacconi}, \& {Valtchanov}}]{wuyts11a}
{Wuyts}, S., {F{\"o}rster Schreiber}, N.~M., {Lutz}, D., {et~al.}
  2011{\natexlab{a}}, \apj, 738, 106

\bibitem[{{Wuyts} {et~al.}(2011{\natexlab{b}}){Wuyts}, {F{\"o}rster Schreiber},
  {van der Wel}, {Magnelli}, {Guo}, {Genzel}, {Lutz}, {Aussel}, {Barro},
  {Berta}, {Cava}, {Graci{\'a}-Carpio}, {Hathi}, {Huang}, {Kocevski},
  {Koekemoer}, {Lee}, {Le Floc'h}, {McGrath}, {Nordon}, {Popesso}, {Pozzi},
  {Riguccini}, {Rodighiero}, {Saintonge}, \& {Tacconi}}]{wuyts11b}
{Wuyts}, S., {F{\"o}rster Schreiber}, N.~M., {van der Wel}, A., {et~al.}
  2011{\natexlab{b}}, \apj, 742, 96

\bibitem[{{Xue} {et~al.}(2011){Xue}, {Luo}, {Brandt}, {Bauer}, {Lehmer},
  {Broos}, {Schneider}, {Alexander}, {Brusa}, {Comastri}, {Fabian}, {Gilli},
  {Hasinger}, {Hornschemeier}, {Koekemoer}, {Liu}, {Mainieri}, {Paolillo},
  {Rafferty}, {Rosati}, {Shemmer}, {Silverman}, {Smail}, {Tozzi}, \&
  {Vignali}}]{chandra4m}
{Xue}, Y.~Q., {Luo}, B., {Brandt}, W.~N., {et~al.} 2011, \apjs, 195, 10

\end{thebibliography}

\clearpage
\label{lastpage}
\end{document}